\documentclass[sigconf]{acmart}

\copyrightyear{2024}
\acmYear{2024}
\setcopyright{acmlicensed}\acmConference[CCS '24]{Proceedings of the 2024 ACM SIGSAC Conference on Computer and Communications Security}{October 14--18, 2024}{Salt Lake City, UT, USA}
\acmBooktitle{Proceedings of the 2024 ACM SIGSAC Conference on Computer and Communications Security (CCS '24), October 14--18, 2024, Salt Lake City, UT, USA}
\acmDOI{10.1145/3658644.3670272}
\acmISBN{979-8-4007-0636-3/24/10}

\AtBeginDocument{%
  \providecommand\BibTeX{{%
    \normalfont B\kern-0.5em{\scshape i\kern-0.25em b}\kern-0.8em\TeX}}}

\usepackage{amsmath, amsfonts, amsthm, amssymb}
\usepackage{xcolor}

\usepackage{graphicx}
\usepackage[ruled, vlined, linesnumbered]{algorithm2e}
\usepackage{caption}
\usepackage{subfigure}
\usepackage{pifont}
\usepackage{makecell,rotating}
\usepackage{algpseudocode}
\usepackage{upgreek}
\usepackage{eucal}
\usepackage{tcolorbox}
\tcbuselibrary{breakable}
\tcbuselibrary{skins}
\usepackage{verbatim}
\usepackage{setspace}
\usepackage{multirow}
\usepackage{booktabs}
\usepackage{tabularx}
\usepackage{bbding,enumitem}
\usepackage{balance}

\newcommand{\ours}{\textsf{Holmes}\xspace}
\newcommand{\rf}{\textsf{RF}\xspace}
\newcommand{\varcnn}{\textsf{Var-CNN}\xspace}
\newcommand{\df}{\textsf{DF}\xspace}
\newcommand{\netclr}{\textsf{NetCLR}\xspace}
\newcommand{\tiktok}{\textsf{Tik-tok}\xspace}
\newcommand{\awf}{\textsf{AWF}\xspace}
\newcommand{\tf}{\textsf{TF}\xspace}
\newcommand{\ares}{\textsf{ARES}\xspace}
\newcommand{\tmwf}{\textsf{TMWF}\xspace}

\newcommand{\pmin}{\textsf{P@min}\xspace}

\newcommand{\first}{\textsf{(i)}\xspace}
\newcommand{\second}{\textsf{(ii)}\xspace}
\newcommand{\third}{\textsf{(iii)}\xspace}

\newcommand\xinhao[1]{{\color{black} #1}}

\begin{document}
\title{Robust and Reliable Early-Stage Website Fingerprinting Attacks via Spatial-Temporal Distribution Analysis}

\author{Xinhao Deng}
\affiliation{
  \institution{INSC \& BNRist, Tsinghua University}
  \city{Beijing}
  \country{China}
}
\email{dengxh23@mails.tsinghua.edu.cn}

\author{Qi Li}
\affiliation{
  \institution{INSC, Tsinghua University}
  \country{}
}
\affiliation{
  \institution{Zhongguancun Laboratory}
  \city{Beijing}
  \country{China}
}
\email{qli01@tsinghua.edu.cn}

\author{Ke Xu}
\affiliation{
  \institution{DCST, Tsinghua University}
  \country{}
}
\affiliation{
  \institution{Zhongguancun Laboratory}
  \city{Beijing}
  \country{China}
}
\email{xuke@tsinghua.edu.cn}

\begin{abstract}

Website Fingerprinting (WF) attacks identify the websites visited by users by performing traffic analysis, compromising user privacy. Particularly, DL-based WF attacks demonstrate impressive attack performance. 
However, the effectiveness of DL-based WF attacks relies on the collected complete and pure traffic during the page loading, which impacts the practicality of these attacks. 
The WF performance is rather low under dynamic network conditions and various WF defenses, particularly when the analyzed traffic is only a small part of the complete traffic. 
In this paper, we propose \ours, a robust and reliable early-stage WF attack. \ours utilizes temporal and spatial distribution analysis of website traffic to effectively identify websites in the early stages of page loading. Specifically, \ours develops adaptive data augmentation based on the temporal distribution of website traffic and utilizes a supervised contrastive learning method to extract the correlations between the early-stage traffic and the pre-collected complete traffic. \ours accurately identifies traffic in the early stages of page loading by computing the correlation of the traffic with the spatial distribution information, which ensures robust and reliable detection according to early-stage traffic. We extensively evaluate \ours using six datasets. Compared to nine existing DL-based WF attacks, \ours improves the F1-score of identifying early-stage traffic by an average of 169.18\%. Furthermore, we replay the traffic of visiting real-world dark web websites. \ours successfully identifies dark web websites when the ratio of page loading on average is only 21.71\%, with an average precision improvement of 169.36\% over the existing WF attacks.

\end{abstract}



\maketitle

\section{Introduction}
\label{sec:intro}

Tor~\cite{dingledine2004tor} is the most popular anonymous communication system, boasting millions of active daily users~\cite{mani2018understanding}.
Tor utilizes various mechanisms, including randomly selected relays and multi-layer encryption, to anonymize user browsing behaviors. 
Unfortunately, Tor is vulnerable to Website Fingerprinting (WF) attacks~\cite{sirinam2018deep,rimmer2018automated, sirinam2019triplet, rahman2019tik, deng2023robust, bahramali2023realistic, jin2023transformer}.
WF attacks utilize Machine Learning (ML) or Deep Learning (DL) models to extract unique traffic patterns of websites and effectively identify the websites visited by Tor users.
In particular, existing DL-based WF attacks demonstrate outstanding attack performance, achieving over 95\% accuracy~\cite{sirinam2018deep,sirinam2019triplet,shen2023rf,deng2023robust}.
WF attacks on Tor traffic are challenging, yet these attacks can also be successfully applied to other privacy-preserving systems~\cite{di2019https, zhan2021website}.

The DL-based WF attacks heavily rely on the collected complete and pure traffic during the page loading for traffic analysis.
In practice, adversaries cannot perceive the entire process of website loading traffic due to mixed background traffic. 
Existing WF attacks apply fixed conditions for traffic collection~\cite{sirinam2018deep, sirinam2019triplet, shen2023rf, rimmer2018automated,deng2023robust}.
These settings do not consider the differences between websites and may compromise the attack performance, e.g., the adversary can only collect partial traffic from slow-loading websites.
Particularly, poor network conditions and WF defenses also prevent the adversary from effectively collecting the complete pure traffic of page loading, leading to a significant decrease in attack performance against certain websites~\cite{juarez2015wtf}.
Our study shows that the robust WF attack (i.e., \df) achieves an average precision of over 91\% for all websites under the WTF-PAD defense. Notably, the lowest precision of fingerprinting is only less than 55\%\footnote{The detailed results can be found in Section~\ref{sec:eval-reliability}.}. 

\begin{figure}[t]
  \centering
  \includegraphics[width=\linewidth]{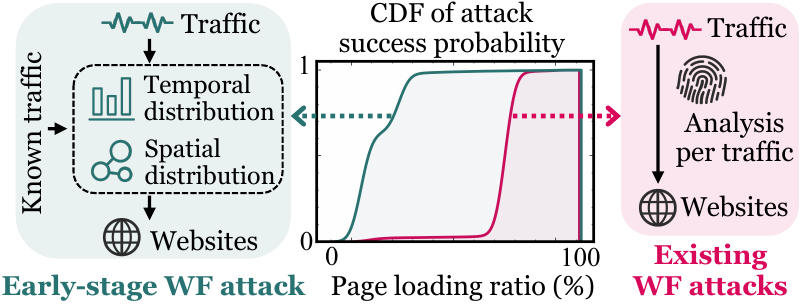}
  \caption{Comparison of the early-stage WF attack with existing WF attacks. The early-stage WF attack can identify websites in the early stage of page loading.}
  \label{fig:cdf_early_attacks}
\end{figure}

To address the limitations of existing DL-based WF attacks, we aim to develop an effective WF attack, i.e., the early-stage WF attack, that only utilizes the traffic generated from the early stage of page loading.  
The early-stage WF attack can identify the visited website during early-stage page loading.
As shown in Figure~\ref{fig:cdf_early_attacks}, compared with existing WF attacks, the early-stage attack does not require waiting for the complete traffic of page loading. However, there are three critical challenges in constructing the early-stage WF attack.
\first Early-stage traffic under dynamic network conditions is prone to traffic misidentification. 
Dynamic network conditions refer to that Tor users may use different paths with different bandwidths and latency across various networks. Under such dynamic network conditions, the patterns of different traffic from the same website vary.
Traffic at the early stages of page loading contains less website information, which varies under dynamic network conditions. 
\second Early-stage WF attacks are more susceptible to various defenses. By padding dummy packets~\cite{juarez2015wtf,gong2020zero}, delaying packets~\cite{cai2014tamaraw,wang2017walkie} or splitting traffic~\cite{de2020trafficsliver}, defenses can significantly impact the effectiveness of WF attacks.
\third The page loading speeds vary significantly across different websites, making it difficult to ensure high precision in detecting early-stage traffic of all websites. Since existing WF attacks based on fixed-setting traffic collection cannot perceive the page loading of websites visited by Tor users,  
the effectiveness is unreliable. Especially, as discussed above, they achieve very low identification precision in detecting the traffic visiting some websites. 

In this paper, we propose \ours\footnote{\ours is a fictional British detective in novels, known for his skill in analyzing the correlations of clues to solve problems earlier than others.}, a robust and reliable early-stage WF attack that can accurately fingerprint traffic visiting different websites according to a small amount of traffic.  
\ours is capable of effectively identifying the early-stage traffic of websites under dynamic network conditions and deployed defenses by correlating the early-stage traffic with the pre-collected complete traffic.
We find that both the early-stage traffic and the complete traffic of the same website exhibit a strong connection of temporal-spatial distribution because they contain the same website information, e.g., the same parts of the website content and elements.  
As illustrated in Figure~\ref{fig:cdf_early_attacks}, \ours achieves early-stage WF attacks by capturing the correlation between the unknown early-stage traffic and the pre-collected complete traffic.

To efficiently capture the correlation between the traffic of different stages of page loading, we design a three-step approach based on temporal-spatial distribution analysis.
First, \ours utilizes an adaptive data augmentation method built on the temporal distribution of traffic features, which augment the traffic at different stages of page loading. 
Second, \ours utilizes supervised contrastive learning to transform traffic features into the low-dimensional embedding space so that traffic at different loading stages is clustered closely in the same embedding space. 
Notably, supervised contrastive learning makes the traffic of the same website closer in the embedding space by learning the correlations of the traffic. 
Third, \ours transforms unknown early-stage traffic into a point in the embedding space and calculates the correlation between the unknown traffic and each website based on the spatial distribution of website traffic. It allows \ours to perform website identification at each short interval of traffic collection. 
The identified results with low confidence will be rejected because the dynamic network conditions or defenses cause insufficient website information in the early-stage traffic. 
\ours automatically continues collecting more packets and analyzing the traffic at the next interval until the website is successfully identified.
Therefore, \ours can ensure adaptive traffic collection and reliable early-stage website identification.

We prototype \ours and conduct extensive performance evaluations using six different datasets, including the Alexa-top websites dataset, dark web websites dataset, and four defense datasets. 
Moreover, we implement nine advanced DL-based WF attacks for comparison with \ours.
Compared to the existing WF attacks, \ours achieves an average improvement of 169.18\% in the F1-score to identify the early-stage traffic.  
Particularly, the experimental results under multiple defenses demonstrate the exceptional robustness of \ours.
Furthermore, we evaluate the performance of \ours under real-world deployment. 
We selected 80 popular dark web websites based on Tor onion services~\cite{wang2023comprehensive} and collected real-world dark web traffic in August 2023 and April 2024. 
\ours achieves a precision of 85.19\% in identifying these real-world dark web websites, with an average page loading ratio of only 21.71\%.

The contributions of our work are three-fold:

\begin{itemize}[leftmargin=*]
    \item We propose \ours, the first robust and reliable early-stage WF attack against Tor traffic, which can fingerprint websites according to a small amount of traffic visiting the websites. 
    \item \ours utilizes feature attribution to analyze the temporal distribution of traffic features, enabling website-adaptive data augmentation. Furthermore, \ours utilizes a supervised contrastive learning method to extract correlations between early-stage traffic and complete traffic and obtain the spatial distribution of websites. By correlating the spatial and temporal distribution, \ours achieves a robust and reliable website identification, which can accurately fingerprint traffic under different network conditions and various defenses. 
    \item We prototype \ours and perform extensive experiments in various settings to demonstrate its performance. We release the source code of \ours\footnote{https://github.com/Xinhao-Deng/Website-Fingerprinting-Library}.
\end{itemize}

The rest of this paper is organized as follows: Section 2 presents the background and the problem statement. Section 3 presents the threat model. 
In Section 4, we present the key observation and overview of \ours.
Section 5 presents the detailed designs. 
In Section 6, we evaluate the performances of \ours. 
In Section 7, we discuss the practicality of \ours and the possible countermeasure against \ours. 
Section 8 and 9 review related works and conclude the paper, respectively. 

\section{Background \& Problem Statement}

\subsection{Background}

Website fingerprinting (WF) attacks identify the websites visited by Tor users by analyzing traffic patterns, such as packet sizes and timing information. 
Previous WF attacks extract fingerprinting features from traffic based on expert knowledge and employ Machine Learning (ML) models to classify these features for website identification~\cite{panchenko2016cumul,wang2014effective,hayes2016k}. 
However, features extracted based on expert knowledge can be easily compromised by defenses~\cite{juarez2015wtf}. 
With the advancement of deep learning (DL), DL-based WF attacks achieve automated feature extraction and significantly enhance performance~\cite{sirinam2018deep,rimmer2018automated,bhat2019varcnn}. 
DL-based WF attacks can effectively identify websites in various real-world scenarios, such as multi-tab browsing~\cite{deng2023robust,xu2018multi,jin2023transformer}, under defenses~\cite{rahman2019tik, shen2023rf}, dynamic network environments~\cite{bahramali2023realistic}, and concept drift~\cite{sirinam2019triplet}. 
However, reliance on the collection of pure traffic throughout the entire page loading hinders the real-world deployment of WF attacks.
\ours achieves early-stage WF attacks by utilizing both the temporal and spatial distributions of website traffic.

Website fingerprinting (WF) defenses aim to undermine the effectiveness of WF attacks. Existing defenses mainly fall into two categories: disturbing traffic and splitting traffic. 
The defenses for disturbing traffic involve padding dummy packets~\cite{juarez2015wtf,gong2020zero}, delaying packets~\cite{wang2017walkie, holland2022regulator}, inserting adversarial perturbations~\cite{nasr2021defeating} and obfuscating traffic~\cite{nasr2021gan3}.
However, the significant overhead of defenses may affect the operation of relay nodes~\cite{cherubin2022online}.
Only a variant of the lightweight defense WTF-PAD has been deployed in the Tor network~\cite{circuitpadding}. 
Traffic splitting defenses involve splitting traffic into multiple paths so that the adversary can only collect a portion of the packets, thereby obscuring the traffic patterns~\cite{de2020trafficsliver}.
We evaluate the robustness of \ours against existing defenses in Section~\ref{sec:robust_evaluation}.

\subsection{Problem Statement}
\label{sec:problem_statement}

\begin{figure}[t]
  \centering
  \includegraphics[width=\linewidth]{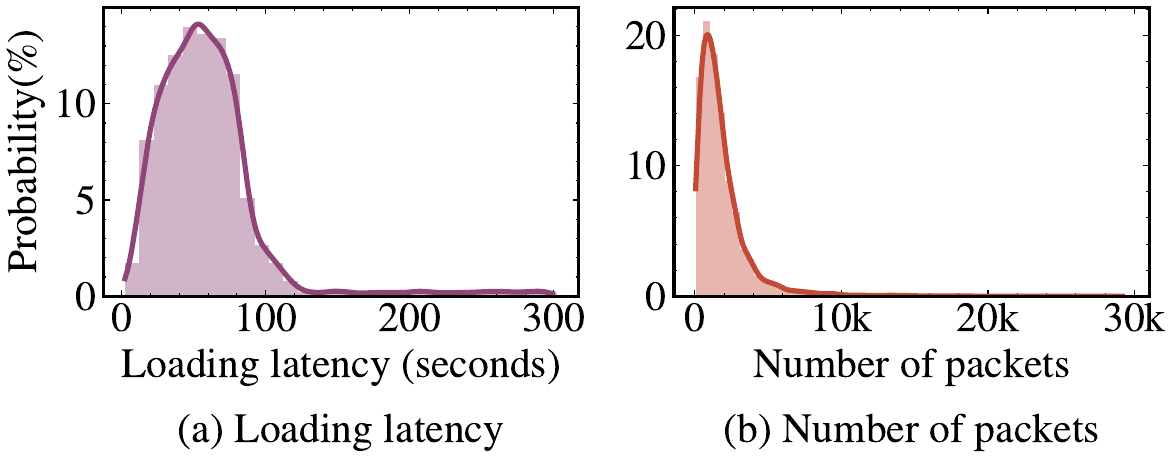}
  \caption{\xinhao{Distribution of page load times and number of packets for Alexa-top 10k websites.}}
  \label{fig:dist_top10k_analysis}
\end{figure}

The goal of this paper is to develop reliable WF attacks (i.e., accurately identifying all websites) based on the traffic in the early stage of page loading. 
Previous WF attacks rely on collecting pure traffic throughout the entire page load process.
However, under dynamic network conditions or defenses, existing attacks cannot effectively collect complete traffic from all websites.
Meanwhile, increasing the traffic collection time incurs more noise from background traffic or defenses, which further impacts the WF performance. 
We analyze the SOTA multi-tab attack \ares~\cite{deng2023robust} and the robust attack \df~\cite{sirinam2018deep}. 
\ares and \df achieve over 90\% average precision in the presence of obfuscated traffic under multi-tab browsing and WTF-PAD defenses, respectively. 
We find that the minimum precision of fingerprinting achieved by \ares and \df is only 42.86\% and 54.11\%, respectively.

Note that, the fixed traffic collection settings required by the existing attacks further undermine the practicality. 
For example, the \df attack sets a traffic collection time of 120 seconds and an input length of 5000. The input is the direction sequence of packets, which is either truncated or zero-padded.
However, different websites exhibit significant variations in page loading latency and the number of generated packets, and such fixed settings cannot guarantee reliable identification of all websites. 
Figure~\ref{fig:dist_top10k_analysis} illustrates the distribution of page loading latency and the number of packets for the Alexa-top 10k websites. 
We observe that the page loading latency of 5.04\% of websites exceeds 120 seconds or the packet count is over 5000, making it difficult for the existing attacks to collect pure traffic with sufficient website information.  
Moreover, we find that over 58.17\% of websites require less than 60 seconds or fewer than 2500 packets for page loading.  
When these websites finish loading, existing attacks continue collecting noise packets, which further degrades the performance of attacks. 

To address the issues above and achieve effective WF attacks at the early stage of page loading, we develop \ours to achieve the following goals.
\first Reliability. \ours utilizes traffic collected from the early stages of page loading to achieve high identification precision across all websites. \second Adaptivity. For traffic from various websites, \ours dynamically performs an attack during each time interval of the traffic collection. \ours should adaptively stop traffic collection once enough website information is obtained, and accurately identify traffic.
\third Robustness. \ours should maintain robust performance under various WF defenses.

In a nutshell, \ours aims to achieve robust and reliable early-stage WF attacks, effectively identifying each website during the
early stages of page loading. 
Compared to previous attacks, \ours may be more practical in the real world, with applications such as early detection and prevention of dark web crimes.
\section{Threat Model}
\label{sec:threat_model}

\begin{figure}[t]
  \centering
  \includegraphics[width=0.95\linewidth]{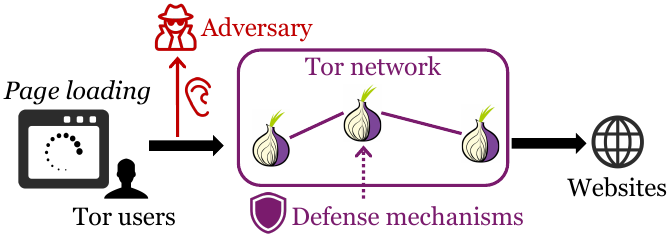}
  \caption{The threat model of \ours.}
  \label{fig:threat-model}
\end{figure}

This paper aims to develop an early-stage website fingerprinting attack that can identify websites visited by Tor users based on the traffic in the early stages of page loading. 
In particular, early-stage WF attacks can identify websites while the Tor user is still waiting for the page to fully load.
In Figure~\ref{fig:threat-model}, we show the threat model of our early-stage WF attack.
Similar with previous works~\cite{hayes2016k,panchenko2016cumul,sirinam2018deep,rimmer2018automated, sirinam2019triplet, rahman2019tik, deng2023robust, bahramali2023realistic, jin2023transformer},
we consider a local and passive adversary for Tor, such as network administrators, Internet Service Providers (ISPs), and Autonomous Systems (AS). 
The adversary can only collect packets without the capability to decrypt packets.
Specifically, a passive adversary is unable to detect the end of a webpage loading, and can only configure fixed conditions for traffic collection~\cite{rimmer2018automated, sirinam2018deep, deng2023robust}.
Furthermore, we consider real-world scenarios with defenses. 
On-path Tor relay nodes can be deployed with defenses, such as padding dummy packets and delaying packets.

Similar to existing attacks~\cite{sirinam2018deep,rimmer2018automated,sirinam2019triplet}, we consider closed-world and open-world scenarios. 
The closed-world scenario assumes that Tor users only visit a limited number of websites.
Therefore, the adversary can collect the traffic from all websites in advance in the closed-world scenario.
In the open-world scenario, clients can browse arbitrary websites, and the adversary can only collect traffic from a small subset of websites. Therefore, Tor users might browse unmonitored websites unknown to the adversary in the open-world scenario.

\section{Design of \ours}
\label{sec:design_overview}

In this section, we present the key observation for our design and propose a robust and reliable early-stage WF attack. 

\subsection{Key Observation}
\begin{figure}[t]
  \centering
  \includegraphics[width=0.9\linewidth]{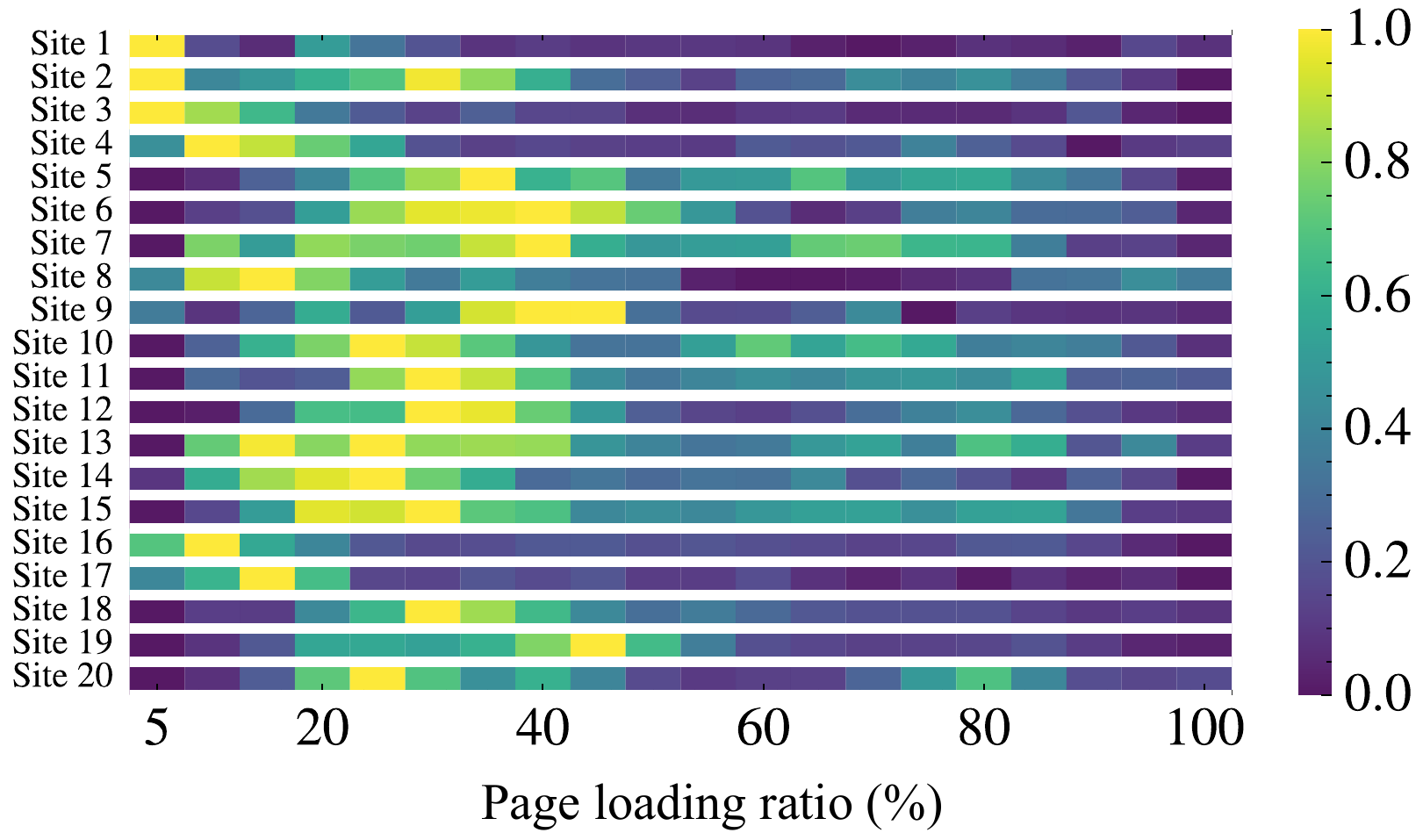}
  \caption{Visualization of temporal distribution based on feature attribution method SHAP~\cite{lundberg2017unified}.}
  \label{fig:shap_analysis}
\end{figure}

As discussed in Section~\ref{sec:problem_statement}, identifying websites by analyzing single early-stage traffic is challenging due to dynamic network conditions and deployed defenses.
Particularly, the loaded content during the same loading interval varies under different network conditions.
However, we observe a strong correlation between the early-stage traffic and the pre-collected complete traffic of the same website, both of which invariably contain the same website information, including parts of the website content and elements.

Figure~\ref{fig:shap_analysis} illustrates the distribution of website information across different stages of page loading, i.e., the temporal distribution of the website features. 
For simplicity without losing generality, we randomly select 20 websites from the Alexa-top 95 websites.
We cannot directly analyze the website information corresponding to the encrypted packets. 
Thus, we measure the importance of traffic features for each page loading stage based on the feature attribution method, i.e., SHAP~\cite{lundberg2017unified}.
The importance of traffic features refers to their contribution to website identification. 
The more website information contained in the page loading stage, the more important the corresponding traffic features.
We observe that the early-stage traffic of all websites shares similar sufficient website information with the complete traffic.
Therefore, it is possible for us to achieve accurate early-stage traffic fingerprinting by analyzing the correlation between the early-stage traffic and the pre-collected complete traffic. 

\subsection{Overview of \ours}
\begin{figure*}[t]
  \centering
  \includegraphics[width=\linewidth]{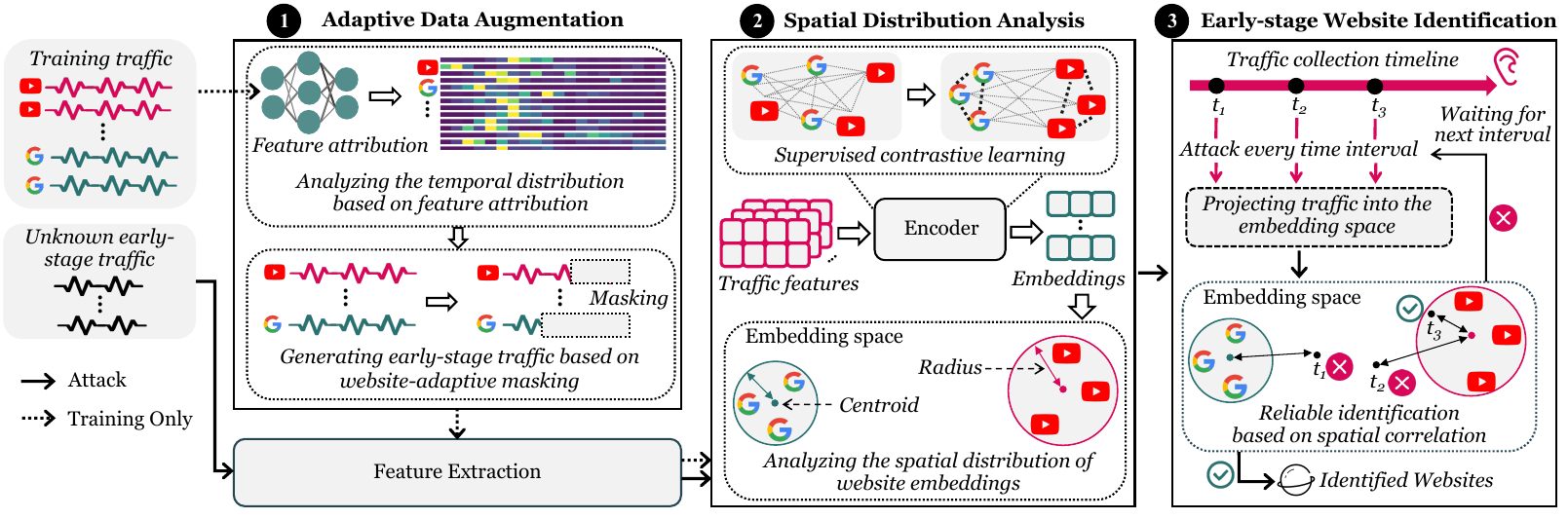}
  \caption{The overview of \ours.}
  \label{fig:overview}
\end{figure*}

In this paper, we propose \ours that exploits the correlations between the early-stage traffic and the pre-collected complete traffic to achieve early-stage WF attacks.
Particularly, \ours captures the spatial and temporal distribution of different websites so that it can accurately fingerprint the traffic according to a small amount of the traffic visiting the websites, even under varied network conditions and WF defenses. 
\ours first performs data augmentation based on the unique temporal distribution of traffic features for each website, which generates  
early-stage traffic that contains sufficient website information.
Second, \ours utilizes Supervised Contrastive Learning (SCL)~\cite{khosla2020supervised} to transform traffic features into a low-dimensional embedding space, where each flow of traffic corresponds to a point in the space.
SCL extracts the correlation between early-stage and complete traffic of the same website by clustering the points of early-stage and complete traffic in the embedding space.
Finally, \ours projects unknown early-stage traffic into the embedding space and calculates its correlation with each website based on the spatial distribution of website traffic in the embedding space.
Note that, to avoid misidentification of early-stage traffic containing only connection information, \ours rejects results of identifying early-stage traffic with low correlations to all websites.
Therefore, \ours performs attacks at each short time interval of traffic collection until the corresponding website is identified with high confidence, thus enabling adaptive traffic collection and reliable identification for each website.

Figure~\ref{fig:overview} illustrates the overview of \ours. 
\ours consists of three modules designed to construct robust and reliable early-stage WF attacks, including adaptive data augmentation, spatial distribution analysis, and early-stage website identification.

\noindent \textbf{Adaptive Data Augmentation.} 
The adaptive data augmentation module generates early-stage traffic by masking the tail of complete traffic during the training phase, which ensures that early-stage traffic contains sufficient website information based on the unique temporal distribution of the website. 
\ours employs the feature attribution method, i.e.,  SHAP~\cite{lundberg2017unified}, to analyze the temporal distribution of the website traffic. It aggregates the feature attribution results of multiple traffic associated with the same websites to obtain the feature importance distribution of the website.
\ours leverages the temporal distribution of websites to apply tail masking of various lengths for the traffic of different websites so that it can adaptively generate early-stage traffic containing sufficient website information for each website.
The details of this module will be described in Section~\ref{sec:5-design-temporal}.

\noindent \textbf{Spatial Distribution Analysis.} 
The spatial distribution analysis module utilizes supervised contrastive learning to transform traffic features and computes the spatial distribution of websites according to the new feature space.
To effectively extract the correlation between early-stage traffic and complete traffic, \ours utilizes an encoder built on supervised contrastive learning to transform traffic features into low-dimensional embedding features, ensuring that the embedding features corresponding to the early-stage and complete traffic of the same website are similar.
The embedding features of traffic are viewed as points in the embedding space, where points corresponding to early-stage and complete traffic with similar embedding features will be clustered together in this space.
\ours analyzes the spatial distribution of each website's traffic in the embedding space, calculating the centroid and radius of each website to support early-stage website identification. 
We will describe this module in Section~\ref{sec:5-design-spatial}.

\noindent \textbf{Early-Stage Website Identification.} 
The early-stage website identification module adaptively collects traffic according to the spatial distribution of websites and achieves reliable website identification.
Since the adversary cannot perceive the page loading progress associated with unknown website traffic, \ours conducts a WF attack during each traffic collection interval.
During each interval, \ours projects the unknown early-stage traffic into the embedding space and then calculates the distance between the point corresponding to the unknown traffic and the centroid of each website. Since different websites have unique distribution densities in the embedding space, i.e. radii, 
we can obtain the correlation between unknown traffic and each website by comparing the distances and radii of websites. 
If the distance between the centroid of a website and the unknown traffic is less than the radius of the website, 
the traffic is successfully identified and traffic collection ends.  Otherwise, \ours will continue collecting traffic and analyze the traffic at the next time interval.
We will present the details of early-stage website identification in Section~\ref{sec:5-design-identification}.
\section{Design Details}
\label{sec:design_details}

In this section, we present the design details of \ours, including the adaptive data augmentation module, the spatial distribution analysis module, and the early-stage website identification module.

\subsection{Adaptive Data Augmentation}
\label{sec:5-design-temporal}

The Adaptive Data Augmentation module generates traffic at different stages of page loading based on masked tail traffic, thereby facilitating the analysis of the correlation between early-stage traffic and complete traffic. 
However, randomly generated early-stage traffic may not contain sufficient website information. The reason is that due to network dynamic conditions and defenses, randomly generated early-stage traffic may only contain connection information and dummy packets.
Furthermore, differences in website loading speed can also affect the correlation between the generated early-stage traffic and the complete traffic.
To achieve website-adaptive data augmentation, \ours utilizes the feature attribution method to analyze the temporal distribution of traffic features, ensuring that the generated early-stage traffic is correlated with the complete traffic of the same website.

\noindent \textbf{Temporal Distribution Analysis.} \ours analyzes the temporal distribution by profiling the feature importance, which is challenging for two reasons: 
\first Packets are encrypted in multiple layers by Tor, making it difficult to analyze their importance. 
\second In dynamic network environments or under traffic obfuscation by defenses, the positions of important packets may change.

To address these challenges, we extend the feature attribution method SHapley Additive exPlanations (SHAP)~\cite{lundberg2017unified} to analyze the feature importance distribution at different stages of page loading.
SHAP calculates the marginal contribution of each feature by generating combinations of all features.
It is based on Shapley values, a concept from cooperative game theory, which ensures a fair distribution of the contribution among the features. SHAP provides local explanations showing how much each feature in a specific instance contributes to the model's output, as well as global insights about the overall model behavior.

The advantages of SHAP over other feature attribution methods include \first Accuracy. SHAP calculates all feature combinations, which enables effective analysis of the relationships among features in encrypted traffic, resulting in more accurate attribution outcomes. \second Consistency. SHAP provides consistent feature attribution results for multiple traffic to the same website. Therefore, \ours can aggregate the feature attribution results of multiple traffic to obtain a website-level distribution of feature importance.

Let $U=\{f_1,f_2,\dots,f_n\}$ represent the feature set of traffic, where $n$ is the number of features.
\ours divides the page loading time into $n$ equal time intervals and counts the number of incoming and outgoing packets in each interval as traffic features, where $f_i$ represents the feature of the i-th interval.
\ours calculates the importance of the i-th feature $f_i$ based on the difference in the expected model output when conditioning on the feature $f_i$.
To miner the dependencies among traffic features, \ours generates all feature combinations excluding the feature $f_i$ to calculate the marginal contribution of the feature $f_i$. Specifically, the importance of the i-th feature $\phi_i$ can be computed as follows:

\begin{equation}
\phi_i = \sum_{S \subseteq U \backslash \{f_i\}} \frac{\vert S \vert! \cdot (n-|S|-1)!}{n!} \cdot (\mathsf{O}(S \cup \{f_i\}) - \mathsf{O}(S)),
\label{equ:shap}
\end{equation}
where $S$ is a feature subset excluding $f_i$. $\mathsf{O}(S \cup \{f_i\})$ and $\mathsf{O}(S)$ represent the expected outputs of the model when feature $f_i$ is present and absent, respectively. The weight of the set $S$ is the frequency of occurrence among all possible feature combinations. 
Subsets of varying sizes are balanced in terms of weight to ensure that the contributions of each feature can be fairly assessed. Due to the high computational cost of Equation~\ref{equ:shap}, we employ the DeepLIFT algorithm~\cite{shrikumar2017learning} for approximation to expedite the calculation.

We select the SOTA WF attack \rf~\cite{shen2023rf} as the target model for feature profiling.
For each website, we randomly select 10 traffic. 
We calculate the importance of features corresponding to different loading stages of the website and represent the temporal distribution of each website using the average temporal distribution of the traffic.

\begin{figure}[t]
  \centering
  \subfigure[Calculation of effective loading ranges.]{
    \label{fig:augmentation_1}
    \includegraphics[width=0.40\linewidth]{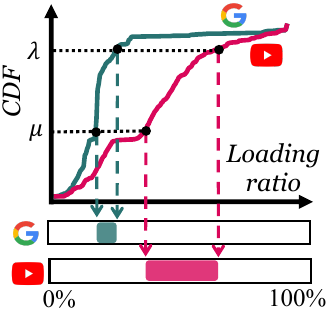}
  }
  \subfigure[Generation of early-stage traffic.]{
    \label{fig:augmentation_2}
    \includegraphics[width=0.46\linewidth]{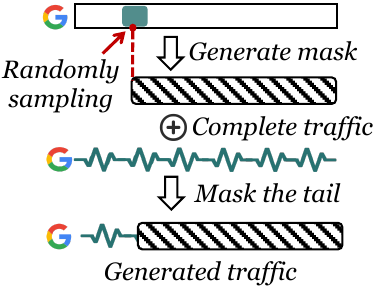}
  }
  \caption{Adaptive data augmentation of \ours. (a) \ours calculates the effective loading ranges of websites based on the temporal distribution of websites. (b) \ours randomly samples the start of the mask based on the effective loading ranges of websites and generates early-stage traffic by masking traffic tails.}
  \label{fig:data_augmentation}
\end{figure}

\noindent \textbf{Mask-based Data Augmentation.} 
Data augmentation is a machine learning technique that enhances the diversity of training data by artificially modifying samples to improve model performance~\cite{bahramali2023realistic}.
\ours achieves the data augmentation by masking the tail of the traffic.
However, the setting of mask proportion is challenging. 
A prolonged mask results in early-stage traffic lacking information related to the website, whereas a too-brief mask requires the adversary to spend a lot of time collecting enough packets.
To address the above challenges, \ours employs website-adaptive data augmentation based on the temporal distribution of websites.

In Figure~\ref{fig:data_augmentation}, we show the details of the data augmentation.
\ours initially calculates the effective loading ranges of websites. 
When the page loading ratio of a website reaches the effective loading range, the early-stage traffic contains enough website information to be correlated with the complete traffic. 
As shown in Figure~\ref{fig:augmentation_1}, \ours generates the cumulative distribution of feature importance for all websites. \ours sets two parameters, $\lambda$ and $\mu$, representing the upper and lower bounds of the cumulative feature importance corresponding to the effective loading proportions of websites.
Based on the parameters $\lambda$ and $\mu$, \ours can calculate the effective loading range for each website.

To ensure the correlation between the generated early-stage traffic and the complete traffic, \ours adaptively enhances the traffic for each website, making the generated traffic originate from the effective loading range of the corresponding website. 
Let $R = \{(s_1, t_1), (s_2, t_2), \dots, (s_m, t_m)\}$ represent the effective loading ranges for $m$ monitored websites, where the effective loading range for the i-th website is from $s_i$ to $t_i$. 
In Figure~\ref{fig:augmentation_2}, we show the details of early-stage traffic generation. 
For the traffic of the i-th website, \ours randomly samples an integer $\boldsymbol{l}$ from $s_i$ to $t_i$, then masks the tail of traffic from the loading ratio $\boldsymbol{l}$ to the entire page loading.
We select the starting point of the mask randomly within an effective range, ensuring that the generated traffic belongs to the early stages of page loading and contains adequate website information.

\begin{equation}
\boldsymbol{l} \sim \texttt{Uniform}[s_i, t_i].
\end{equation}

\ours performs data augmentation on each traffic $\alpha$ times. 
The higher the value of $\alpha$, the more early-stage traffic is generated. However, excessive generation of early-stage traffic can lead to significant time overhead of model training.

\subsection{Spatial Distribution Analysis}
\label{sec:5-design-spatial}

\begin{figure}[t]
  \centering
  \includegraphics[width=\linewidth]{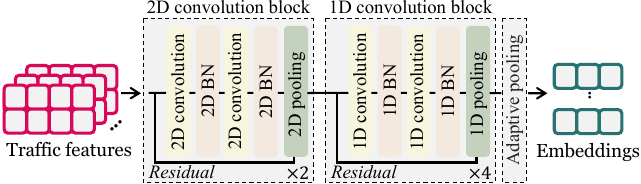}
  \caption{The Encoder of \ours.}
  \label{fig:design_encoder}
\end{figure}

Utilizing the early-stage traffic generated by the temporal distribution analysis module, the spatial distribution analysis module extracts the correlation between early-stage and complete traffic. 
Specifically, \ours builds an Encoder based on Supervised Contrastive Learning (SCL)~\cite{khosla2020supervised} to extract common features of early-stage and complete traffic, generating low-dimensional embeddings that are spatially proximate. 
Then \ours analyzes the spatial distribution of websites using the Median Absolute Deviation (MAD)~\cite{leys2013detecting}.

\noindent \textbf{Traffic Embedding Based on SCL.} 
To address the challenges posed by network jitter and defenses in the real world on the analysis of early-stage traffic, \ours employs Supervised Contrastive Learning (SCL) for traffic embedding. 
The generated embeddings encompass robust features of the traffic, enabling traffic from different loading stages of the same website to aggregate in the embedding space.
Note that, \ours addresses the limitation of the clustering methods, i.e., they cannot effectively aggregate original high-dimensional features due to the “curse of dimensionality”~\cite{zimek2012survey}.

\ours initially extracts raw features from traffic, serving as the input for generating embeddings. 
We use the Traffic Aggregation Features (TAF) as the raw features. TAF is an extension of the Traffic Aggregation Matrix (TAM)~\cite{shen2023rf} that effectively represents aggregated traffic information.
We set up $\rho$ non-overlapping time windows of equal length. The length of the time window is $\theta$. 
For the i-th time window, we calculate three types of aggregated features: \first the number of incoming and outgoing packets. \second the number of incoming and outgoing bursts. \third the average size of incoming and outgoing bursts.
Therefore, we can aggregate statistical information from multiple time windows as the initial feature of the traffic.

We use the Convolutional Neural Network (CNN) as the Encoder network for traffic embedding. CNN is applied by previous attacks~\cite{sirinam2018deep,rimmer2018automated,sirinam2019triplet, shen2023rf, bhat2019varcnn, rahman2019tik} and proved to be effective in extracting key patterns of traffic associated with the website.
Let $\texttt{Enc}(\cdot)$ denote the encoder network, and we can obtain the embedding $\boldsymbol{z}$ of the traffic with the raw feature $\boldsymbol{x}$ based on the Encoder.

\begin{equation}
\boldsymbol{z} = \texttt{Enc}(\boldsymbol{x}).
\end{equation}

We show the details of the Encoder in Figure~\ref{fig:design_encoder}. To effectively extract the correlation of the website traffic at different loading stages, we utilize convolution with a greater number of channels and a deeper network architecture compared to previous attacks~\cite{sirinam2018deep, shen2023rf}. 
Since the input is two-dimensional features, \ours uses two 2D convolution blocks to extract high-dimensional information. Then \ours fusions information of packets with different directions through the 2D pooling layer, transforming the two-dimensional features into one-dimensional features. 
Subsequently, four 1D convolution blocks are utilized to extract traffic patterns related to the website. Finally, \ours employs an adaptive pooling layer to generate embeddings of traffic. 

Furthermore, we employ two complementary methods. First, residual connections are utilized, which involve transmitting intermediate outputs from lower to higher layers via skip connections, thereby reducing the issue of gradient vanishing. Second, multiple dropout layers are used, where a subset of units, including their associated connections, are randomly omitted from the network during the training, thus mitigating overfitting.

The performance of the Encoder depends on effective model training. 
The Encoder aims to extract various correlations in traffic, including
\first The correlation between the traffic at different loading stages of the same website. \second The correlation between the traffic of the same website where the traffic patterns change due to network dynamics or defenses.
Contrastive learning and metric learning can learn the correlation between samples. 
However, both contrastive learning and metric learning consider only one type of correlation that exists in the samples.
To effectively extract multiple types of correlations existing in the samples, \ours applies SCL to train the Encoder. 
SCL combines the advantages of supervised and contrastive learning. 
Specifically, \ours randomly selects one traffic as the anchor. Then, \ours selects all traffic of the anchor's corresponding website as positive samples and traffic of other websites as negative samples. \ours repeats this process multiple times to ensure that the selected anchors include multiple traffic for all websites.

SCL can learn various correlations between the anchor and the positive samples, ensuring that in the generated embedding space, the distance between the anchor and positive samples is close, while the distance between the anchor and negative samples is far. 
Formally, for the i-th traffic $\boldsymbol{x}_i$ with embedding $\boldsymbol{z}_i$, we can calculate its loss by SCL:

\begin{equation}
\mathcal{L}_i = - \frac{1}{\vert \mathrm{P}(i) \vert} \sum_{p \in \mathrm{P}(i)} \texttt{log} \frac{\texttt{exp}(\mathrm{z}_i \cdot \mathrm{z}_p / \gamma) }{\sum_{n \in \mathrm{N}(i)} \texttt{exp}(\mathrm{z}_i \cdot \mathrm{z}_n / \gamma)},
\label{equ:loss}
\end{equation}
where $\mathrm{P}(i)$, $\mathrm{N}(i)$ are the set of the index of all positive samples and negative samples of the i-th traffic, respectively. 
For the embedding of anchor $z_i$, we calculate the similarity with each positive sample embedding $z_p$ and compare it with similarities between the anchor and all negative samples. 
In particular, $\gamma$ is temperature, a hyperparameter that controls the distance of traffic $\boldsymbol{x}_i$ from the most similar negative sample. The smaller the temperature $\gamma$, the greater the differentiation from the negative samples, but it tends to affect the similarity to the positive samples. 
Through Equation~\ref{equ:loss}, we can effectively train the Encoder and extract the correlations of website traffic at different loading stages.

\begin{algorithm}[t]
    \small
    \SetAlgoLined
    \SetKwProg{Fn}{Function}{:}{}
    \SetKwFunction{Main}{Main}
	\caption{Website Profiling} 
	\label{alg:website-profiling}

        \KwIn{ 
        \\
        \quad ${W}$: all websites.  \\
         \quad ${z}$: the embeddings of all websites. \\
         }
         \KwOut{
            \\
        \quad ${c}$: the centroids of all websites. \\
        \quad ${r}$: the radii of all websites. \\ 
         }

	

           \For{$w \in W$}{
                $c_w = \mathtt{Mean}(z^w)$ \Comment{\textit{Calculate the centroid of website} $w$} \\
                \For{$ z^w_i \in z^w $}{
                    $d^w_i = 1 - \mathtt{cosine\_similarity}(c_w, z^w_i)$ \\
                }
                $M^w = \mathtt{Median}(d^w)$ \Comment{\textit{Calculate the median}} \\
                $r_w = \mathtt{Median} \{\vert d^w_i - M^w \vert$ \} \Comment{\textit{Calculate the radius}}\\
            }
            \For{$ w_i, w_j \in W$}{
                     ${d} = 1 - \mathtt{cosine\_similarity}({c}_i, {c}_j)$ \\
                    \If{${{r}}_{{i}} + {{r}}_{{j}} \geq {d} $}{
                        \Comment{\textit{Tuning the radius}} \\
                        ${{r}}_{{i}} = {{r}}_{{i}} - \frac{{{r}}_{{i}}}{{{r}}_{{i}} + {{r}}_{{j}}} \cdot ({{r}}_{{i}} + {{r}}_{{j}} - {d})$ \\
                        ${{r}}_{{j}} = {{r}}_{{j}} -   \frac{{{r}}_{{j}}}{{{r}}_{{i}} + {{r}}_{{j}}} \cdot ({{r}}_{{i}} + {{r}}_{{j}} - {d})$ \\
                    }
                }
                
            \Return ${\mathit{c}}$, ${\mathit{r}}$        
\end{algorithm}

\noindent \textbf{Spatial Distribution Based on MAD.} 
\ours aims to achieve reliable early-stage website identification. However, the early-stage traffic contains little website information and is prone to misidentification under the interference of network dynamics and defenses. 
\ours addresses the challenge by utilizing the spatial distribution of website traffic. 
Traffic from different websites has different positions and levels of tightness in the embedding space. 
\ours calculates the centroid and radius for each website, representing the position and level of tightness of the website traffic, respectively. 
By leveraging the centroid and radius information of websites, \ours can reject low-confidence identifications of unknown traffic. 
We will detail how to utilize the centroid and radius of websites for reliable early-stage website identification in Section~\ref{sec:5-design-identification}.

In Algorithm~\ref{alg:website-profiling}, we show the pseudocode for website profiling. 
Suppose there are $m$ websites. 
\ours sequentially calculates the centroid and radius for each website.
For the website $w$, \ours generates the embeddings for all traffic of the website $w$.
Let $z_i^w$ represent the embedding of the i-th traffic of the website $w$.
\ours calculates the centroid of website $w$ by averaging all embeddings across each dimension (line 2).
Then \ours calculates the distance between each traffic embedding and the centroid of the website $w$ using cosine similarity (lines 3-4). 
We select cosine similarity because matrix operations can accelerate multiple cosine similarity calculations.
Finally, we use a distribution estimation algorithm, Mean Absolute Deviation (MAD)~\cite{leys2013detecting} to generate the radius for the website $w$ (lines 6-7). 
MAD calculates the median of absolute deviations, where absolute deviation refers to the absolute value of the difference between each data and the median of all data. 

Based on the centroid and radius of each website, the spatial distribution of each website in the embedding space forms a sphere.
\ours utilizes supervised contrastive learning to separate the centroids of different websites in the embedding space. However, we observe a 0.01\% probability of overlap between the spheres corresponding to the two websites in our study. 
This occurs because the centroids of websites with similar types or content are closer to each other. 
Therefore, \ours further examines the distances between the centroids of different websites and their corresponding radii. 
For two websites $w_i$ and $w_j$, if the distance between $\mathrm{c}_i$ and $\mathrm{c}_j$ is less than the sum of the radii, we proportionally reduce the radii of website $\mathrm{w}_i$ and website $\mathrm{w}_j$.
Finally, the spheres corresponding to each website in the embedding space are non-overlapping, which facilitates the early-stage website identification of \ours.

\subsection{Early-Stage Website Identification}
\label{sec:5-design-identification}

The early-stage website identification module leverages the correlations between different loading stages of website traffic to achieve robust and reliable identification of early-stage traffic. 
To achieve early-stage website identification, \ours attempts website identification at each fixed time interval.
The challenge faced by \ours is ensuring high confidence in website identification to avoid misidentification of early-stage traffic. 
To address the above challenge, \ours calculates the correlation between unknown traffic and monitored websites based on the position of unknown traffic in the feature space and the spatial distribution of monitored websites.
\ours rejects the identification of early-stage traffic with low correlation to all monitored websites and continues to collect more packets.

\begin{algorithm}[t]
    \small
    \SetAlgoLined
    \SetKwProg{Fn}{Function}{:}{}
    \SetKwFunction{Main}{Main}
	\caption{Early-stage Website Identification} 
	\label{alg:website-identification}

\KwIn{ 
        \\
         \quad ${\tau}$: the time interval. \\
        \quad ${\sigma}$: the maximum traffic collection time. \\
        \quad ${W}$: all monitored websites.  \\
        \quad ${c}$: the centroids of all monitored websites. \\
        \quad ${r}$: the radii of all monitored websites. \\
        \xinhao{\quad ${\hat{w}}$: the unmonitored website.} \\
        \xinhao{\quad $\epsilon$: threshold for concept drift detection.}
         }
         \KwOut{
            \\
        \quad ${\mathtt{res}}$: the identification result.
         }

            \xinhao{$\mathtt{res} = \hat{w}$} \\ 
            $\mathtt{count} = 0$ \\
           \While{$\mathtt{True}$}{
                $\mathtt{time.sleep}(\tau)$ \Comment{\textit{Wait time interval} $\tau$} \\
                $\mathtt{count} = \mathtt{count} + \tau$ \\
                $\mathbf{x} = \mathtt{getTraffic}()$ \Comment{\textit{Get the current collected traffic}} \\
                $\mathbf{z} = \mathtt{Encoder}(\mathbf{x})$ \\
                \For{$w \in W$}{
                    $d = 1 - \mathtt{cosine\_similarity}(c_w, \mathbf{z})$ \\
                    \If{$d \leq r_w$}{
                        $\mathtt{res} = w$ \Comment{\textit{Identification success}}\\
                        $\mathtt{break}$ \\
                    }
                }
                \If{\xinhao{($\mathtt{res} \neq \hat{w}$)} $\mathtt{or}$ ($\mathtt{count} > \sigma$)}{
                     $\mathtt{break}$ \Comment{\textit{Exit identification}}\\
                }
           }
           
           \If{\xinhao{$\mathtt{res} == \hat{w}$}}{ 
                \xinhao{$d^{min} = \epsilon$} \\
                \For{$w \in W$}{
                    $d = 1 - \mathtt{cosine\_similarity}(c_w, \mathbf{z})$ \\
                    \If{$d - r_w  < d^{min}$}{
                        $d^{min} = d - r_w$ \\
                        $\mathtt{res} = w$ \\
                    }
                    
                }
            }
            \Return $\mathtt{res}$       
\end{algorithm}

In Algorithm~\ref{alg:website-identification}, we show the pseudocode for early-stage website identification.
At every time interval, \ours first projects the unknown early-stage traffic into the embedded space (lines 6-7) and calculates the distance between the unknown traffic and the centroids of all monitored websites (lines 8-9). 
If the distance between the unknown traffic and a website's centroid is less than the radius of the website, \ours successfully identifies the traffic (lines 10-12). Otherwise, \ours continues to collect traffic and waits for the next time interval.

However, not all early-stage traffic can be guaranteed to be identified. 
Changes in the content of monitored websites can lead to variations in traffic patterns (i.e., concept drift). Furthermore, \ours is unable to detect early-stage traffic from unmonitored websites. \ours sets a maximum traffic collection time $\sigma$. 
After collecting traffic for $\sigma$ seconds, \ours will detect whether the unknown traffic is due to concept drift or originates from unmonitored websites (lines 19-28).
A key insight is that the distance between a website's concept drift traffic and its centroid should be slightly greater than the website's radius, yet much smaller than the distance between unmonitored website traffic and the website's centroid.
Therefore, we set a predefined threshold $\epsilon$.
If the difference between the distance of the unknown traffic from the website's centroid $d$ and the website's radius $r_w$ is less than $\epsilon$, then the unknown traffic is identified as a concept drift sample of the website (lines 21-25).
Furthermore, we define a variable $d^{min}$ to represent the smallest difference between $d$ and $r_w$ among all websites, with $d^{min}$ initially set to the threshold $\epsilon$ (line 20). 
If the unknown traffic meets the concept drift detection criteria for multiple monitored websites, we identify the traffic as the website with the highest correlation, which is the website corresponding to $d^{min}$.
In particular, we set the threshold for concept drift detection $\epsilon$ to infinity in the closed-world scenario. The reason is that in the closed-world scenario, Tor users only visit monitored websites, eliminating the need to identify traffic from unmonitored websites.
\section{Performance Evaluation}
\label{sec:evaluation-total}

In this section, we evaluate \ours with public datasets and real-world datasets. We compare the performance of \ours with the state-of-the-art WF attacks.

\begin{table}[t]
\small
\centering
\caption{Parameter settings 
in our evaluation}
\label{tab:parameters}
\begin{tabular}{ccc}
\toprule
\textbf{Group} & \textbf{parameters} & \textbf{Value} \\ \midrule
\multirow{3}{*}{Data Augmentation} 
 & Lower bound of CDF $\mu$ & {0.3} \\
 & Upper bound of CDF $\lambda$ & {0.6} \\
& Number of augmentation $\alpha$ & {2} \\ \midrule
\multirow{4}{*}{Spatial Analysis}
 & Number of time windows $\rho$ & 2000 \\
 & Length of time windows $\theta$ & 80 ms\\
 & Embedding size $\eta$ & 128 \\
 & Temperature $\gamma$ & 0.1 \\ \midrule
\multirow{3}{*}{Website Identification} 
 & Time interval $\tau$ & 120 ms \\
 & Maximum collection time $\sigma$ & 80 s \\
& \xinhao{Threshold for concept drift $\epsilon$} & \xinhao{0.01} \\
\bottomrule
\end{tabular}
\end{table}

\subsection{Experimental Setup} 
\label{sec:evaluation-setup}

\noindent \textbf{Implementation.} 
We prototype \ours using PyTorch 2.0.1 and Python 3.8 with more than 1,400 lines of code. In particular, we use a single NVIDIA GeForce RTX 4090 GPU for our experiments. 
We show the default parameter values in Table~\ref{tab:parameters}. 
Furthermore, we split the dataset into training, validation, and testing, with an 8:1:1 ratio. 
The parameter tuning and spatial-temporal analysis are performed on the validation dataset to avoid leakage of the testing dataset.

\noindent \textbf{Dataset.}
Our datasets comprise six categories of data, including a dataset of Alexa-top websites, a dataset of dark web websites, and four types of defended datasets.

\begin{itemize}[leftmargin=*]
\item {\textbf{Dataset of Alexa-top websites}}: 
This dataset is from~\cite{sirinam2018deep}, which includes data from both closed-world and open-world scenarios. The closed-world data comprises 95 monitored websites, each with over 1000 traces. In the open-world scenario, there are over 40,000 unmonitored websites, each with only one trace. 
All websites belong to the Alexa-top websites list, which ranks websites based on popularity.

\item {\textbf{Dataset of dark web websites}}: 
Since Alexa-top does not represent the popularity of visits by Tor users, we select 80 of the most popular dark web websites based on the measurement of Tor v3 onion services~\cite{wang2023comprehensive}.
The dataset includes various types of websites, comprising black markets, social networks, and financial services.
These websites use onion services to anonymize servers, requiring more relay nodes and resulting in greater loading latency. 
We utilized 20 servers deployed across three countries to collect traffic in August 2023 and April 2024.
Note that our data collection did not negatively impact the real-world Tor network. 
We only collect traffic from browsing sessions we initiated locally, ensuring our dataset does not include data from other Tor clients.

\item {\textbf{Dataset with WTF-PAD defense}}: The WTF-PAD defense~\cite{juarez2015wtf} disrupts traffic patterns by adaptively padding dummy packets without delaying any packets. The variation of WTF-PAD defense based on circuit-level padding has been deployed in Tor~\cite{circuitpadding}.

\item {\textbf{Dataset with Front defense}}: The Front defense~\cite{gong2020zero} utilizes the Rayleigh distribution to generate the padding times for dummy packets. Similar to the WTF-PAD defense, the time overhead for the Front defense is zero.

\item {\textbf{Dataset with Walkie-Talkie defense}}: Walkie-Talkie~\cite{wang2017walkie} employs a half-duplex communication model and merges original traffic with traffic from randomly selected decoy pages to mislead WF attacks. This defense introduces a mild bandwidth and time overhead.

\item {\textbf{Dataset with TrafficSliver defense}}: The TrafficSliver defense~\cite{de2020trafficsliver} employs a traffic-splitting mechanism that restricts the adversary to collecting only partial packets. We generate the dataset by splitting the traffic into three paths based on the script provided by the authors.

\end{itemize}

WF defenses have been extensively studied~\cite{cai2014tamaraw, gong2022surakav, dyer2012peek, cai2014cs, juarez2015wtf, gong2020zero, de2020trafficsliver, wang2017walkie}, yet some defenses are not practically deployable due to the significant overhead~\cite{mathews2023sok}. 
The latency introduced by defenses may cause out-of-memory errors in Tor relay nodes. 
Therefore, following previous attacks~\cite{shen2023rf,rahman2019tik,sirinam2018deep}, we select four representative defense methods for evaluation: WTF-PAD~\cite{juarez2015wtf}, Front~\cite{gong2020zero}, TrafficSliver~\cite{de2020trafficsliver}, and Walkie-Talkie~\cite{wang2017walkie}.

\noindent \textbf{Baselines.} 
We select 9 state-of-the-art WF attacks as our baselines.

\begin{itemize}[leftmargin=*]
\item {\textbf{\textsf{AWF}}}: \textsf{AWF}~\cite{rimmer2018automated} utilizes CNNs to automatically extract features from packet direction sequences for website identification.

\item {\textbf{\textsf{DF}}}: \textsf{DF}~\cite{sirinam2018deep} proposes more sophisticated CNNs compared to AWF that can effectively undermine WTF-PAD defense.

\item {\textbf{\textsf{Tik-Tok}}}: \textsf{Tik-Tok}~\cite{rahman2019tik} utilizes both direction and timestamp information of packets, which can effectively improve attack performance under defense.

\item {\textbf{\textsf{Var-CNN}}}: \textsf{Var-CNN}~\cite{bhat2019varcnn} designs a more powerful model based on ResNets, which utilizes mechanisms such as dilated convolution to improve attack performance. 

\item {\textbf{\textsf{TF}}}: \textsf{TF}~\cite{sirinam2019triplet} extends the DF model using Triplet networks to achieve the best performance in scenarios with fewer training instances.

\item {\textbf{\textsf{RF}}}: \textsf{RF}~\cite{shen2023rf} extracts a two-dimensional matrix feature named TAM, which has better robustness against defenses.

\item \textbf{\textsf{NetCLR}}: \textsf{NetCLR}~\cite{bahramali2023realistic} integrates data augmentation and self-supervised learning. It introduces three data augmentation methods for traffic bursts to improve the effectiveness of WF attacks in dynamic network environments.

\item \textbf{\textsf{ARES}}: \textsf{ARES}~\cite{deng2023robust} is a robust multi-tab WF attack that integrates multiple Transformer-based classifiers to identify websites within obfuscated traffic. ARES also supports single-tab WF attacks.

\item \textbf{\textsf{TMWF}}: \textsf{TMWF}~\cite{jin2023transformer} applies \textsf{DETR}~\cite{carion2020DTER}, a Transformer-based object detection framework, to achieve multi-tab WF attacks. We set the number of tab queries to 1 to apply TMWF to single-tab WF attacks.

\end{itemize}

To reduce the time overhead of the experiments, the parameters of baselines are all set to their default values.
Note that the baselines may achieve better performance with parameter tuning.

\noindent \textbf{Metrics.} We select 4 metrics that are widely used to evaluate the performance of WF attacks, i.e., Accuracy, Precision, Recall, and F1-score. We calculate the macro average of all websites. Specifically, we can calculate the numbers of true positive instances ($\mathtt{TP}$), false positive instances ($\mathtt{FP}$), true negative instances ($\mathtt{TN}$), and false negative instances ($\mathtt{FN}$) for each website, respectively. These four metrics can be calculated as:

\begin{equation}
    \mathtt{Accuracy} = \frac{\mathtt{TP}+\mathtt{TN}}{\mathtt{TP}+\mathtt{TN}+\mathtt{FP}+\mathtt{FN}}.
\end{equation}

\begin{equation}
    \mathtt{Precision} = \frac{\mathtt{TP}}{\mathtt{TP}+\mathtt{FP}}.
\end{equation}

\begin{equation}
    \mathtt{Recall} = \frac{\mathtt{TP}}{\mathtt{TP}+\mathtt{FN}}.
\end{equation}

\begin{equation}
    \mathtt{F1-score}=\frac{2 \times \mathtt{Precision} \times \mathtt{Recall}}{\mathtt{Precision} + \mathtt{Recall}}.
\end{equation}

The differences in website types and content lead to variations in traffic patterns, and the average precision may obscure the low identification precision of some websites. 
Therefore, we use \pmin to represent the lowest precision across all websites. We can evaluate the reliability of WF attacks by calculating \pmin. 
Furthermore, the base rate fallacy~\cite{juarez2014critical} can lead to an overestimation of the precision in the open-world setting. Following previous attacks~\cite{wang2020high}, we use \texttt{r-precision} for open-world evaluation. Specifically, \texttt{r-precision} assumes that the frequency of visits to unmonitored websites is $r$ times that of monitored websites, hence the sample weight of unmonitored websites is $r$ times that of monitored websites when calculating precision. We set $r$ to 20 in our experiments.

\subsection{Closed-World Evaluation}
\begin{figure}[t]
  \centering
  \includegraphics[width=\linewidth]{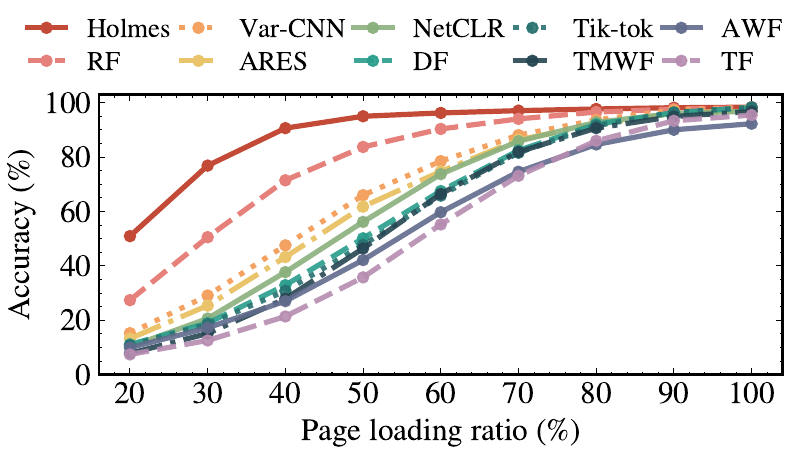}
  \caption{Comparison of WF attacks at different loading stages of websites in the closed-world scenario.}
  \label{fig:early_traffic_acc}
\end{figure}
\begin{table*}[t]
\centering
\footnotesize
\caption{\xinhao{Comparisons with prior arts with the early-stage traffic in the closed-world scenario, where \textbf{P}, \textbf{R}, \textbf{F1} represent Precision (\%), Recall (\%), and F1-score (\%). 
}}
\label{tab:closed-single-tab}
\resizebox{\textwidth}{!}{%
\begin{tabular}{c|ccc|ccc|ccc|ccc|ccc} \toprule

 & \multicolumn{3}{c|}{\textbf{20\% loaded}}
 & \multicolumn{3}{c|}{\textbf{30\% loaded}} 
 & \multicolumn{3}{c|}{\textbf{40\% loaded}} 
& \multicolumn{3}{c|}{\textbf{50\% loaded}} 
 & \multicolumn{3}{c}{\textbf{60\% loaded}} \\ 
 \cmidrule(lr){2-4} 
 \cmidrule(lr){5-7} 
 \cmidrule(lr){8-10} 
 \cmidrule(lr){11-13}  
 \cmidrule(lr){14-16} 
\multirow{-2}{*}{\textbf{\textbf{Attacks}}} & 
\multicolumn{1}{c}{\textbf{P}} & \multicolumn{1}{c}{\textbf{R}} & \multicolumn{1}{c|}{\textbf{F1}} & 
\multicolumn{1}{c}{\textbf{P}} & \multicolumn{1}{c}{\textbf{R}} & \multicolumn{1}{c|}{\textbf{F1}} & 
\multicolumn{1}{c}{\textbf{P}} & \multicolumn{1}{c}{\textbf{R}} & \multicolumn{1}{c|}{\textbf{F1}} & 
\multicolumn{1}{c}{\textbf{P}} & \multicolumn{1}{c}{\textbf{R}} & \multicolumn{1}{c|}{\textbf{F1}} & 
\multicolumn{1}{c}{\textbf{P}} & \multicolumn{1}{c}{\textbf{R}} & \multicolumn{1}{c}{\textbf{F1}} \\  

\midrule 
\textsf{TF} 
& {24.74} & {7.48}  & {8.50} 
& {30.25} & {12.63} & {14.15} 
& {36.55} & {21.44} & {23.31} 
& {48.14} & {35.79} & {37.76} 
& {61.72} & {55.23} & {56.20} 
\\
\textsf{AWF} 
& {28.39} & {9.97}  & {11.75} 
& {33.40} & {17.26} & {19.22} 
& {41.12} & {27.06} &{29.14} 
& {51.16} & {42.15} & {43.61} 
& {63.63} & {59.71} & {59.93}
\\
\xinhao{
\textsf{TMWF}} 
& \xinhao{25.77} & \xinhao{7.87}  & \xinhao{8.27} 
& \xinhao{31.70} & \xinhao{15.19} & \xinhao{16.53} 
& \xinhao{41.15} & \xinhao{27.95} & \xinhao{29.79} 
& \xinhao{56.04} & \xinhao{46.42} & \xinhao{47.78} 
& \xinhao{71.84} & \xinhao{66.44} & \xinhao{67.02}

\\
\textsf{Tik-tok} 
& {37.91} & {10.83} & {11.50} 
& {40.54} & {18.56} & {20.08} 
& {47.00} & {30.82} & {32.89} 
& {59.03} & {47.71} & {49.19} 
& {71.46} & {65.84} & {66.60}
\\
\textsf{DF} 
& {35.28} & {11.19} & {12.62} 
& {40.43} & {19.00} & {21.38} 
& {50.85} & {32.96} & {35.08} 
& {61.52} & {50.17} & {51.68} 
& {72.76} & {67.54} & {68.03} 
\\
\textsf{NetCLR} 
& {32.39} & {10.32} & {11.85} 
& {41.07} & {20.67} & {23.40} 
& {54.19} & {37.72} & {39.96} 
& {65.71} & {56.17} & {57.47} 
& {77.01} & {73.72} & {73.69} 
\\
\xinhao{
\textsf{ARES}} 
& \xinhao{43.06} & \xinhao{13.30} & \xinhao{15.66} 
& \xinhao{51.31} & \xinhao{25.43} & \xinhao{28.70} 
& \xinhao{60.36} & \xinhao{43.28} & \xinhao{45.86} 
& \xinhao{69.77} & \xinhao{61.80} & \xinhao{62.71} 
& \xinhao{77.96} & \xinhao{74.73} & \xinhao{74.57} 

\\
\textsf{Var-CNN} 
& {49.66} & {15.28} & {18.29} 
& {57.66} & {29.12} & {32.85} 
& {65.49} & {47.52} & {50.39} 
& {74.21} & {65.98} & {67.22} 
& {81.64} & {78.51} & {78.69} 
\\
\textsf{RF} 
& {55.51} & {27.44} & {31.27} 
& {67.32} & {50.55} & {53.17} 
& {78.23} & {71.43} & {72.43} 
& {86.22} & {83.70} & {84.06} 
& {91.34} & {90.34} & {90.41} 
\\
\textsf{\textbf{\ours}} 
& \textbf{66.79} & \textbf{50.92} & \textbf{53.45} 
& \textbf{80.22} & \textbf{76.85} & \textbf{76.48} 
& \textbf{91.14} & \textbf{90.64} & \textbf{90.48} 
& \textbf{95.19} & \textbf{95.01} & \textbf{95.00} 
& \textbf{96.40} & \textbf{96.24} & \textbf{96.23} 
\\
\bottomrule
\end{tabular}%
}
\end{table*}

We first evaluate the performance of \ours in the closed-world scenario using the dataset of Alexa-top 95 websites.
To assess the performance of \ours in identifying early-stage website traffic, we generate traffic for different loading stages of websites based on packet timestamps from the testing dataset. 
As shown in Figure~\ref{fig:early_traffic_acc}, \ours achieves optimal attack performance under different page loading ratios.
As the loading progress of websites increases from 20\% to full completion, the Accuracy of \ours in identifying the website gradually improves, rising from 50.94\% to 98.36\%. 
Compared to existing attacks, \ours demonstrates a significant advantage in early-stage traffic analysis. 
For example, when websites are 40\% loaded, \ours achieves an Accuracy of 90.65\%, 
which represents an improvement of 26.84\%, 90.68\%, 109.50\%, 140.32\%, 175.36\%, 194.22\%, 224.68\%, 235.12\%, and 323.60\% over \rf, \varcnn, \ares, \netclr, \df, \tiktok, \tmwf, \awf, and \tf, respectively.
Specifically, \ours exhibits the highest Accuracy for traffic at all loading stages of websites. 
The primary reason is that \ours extracts traffic correlations at different loading stages of websites through spatial-temporal analysis. This correlation enhances the ability of \ours to identify traffic across all loading stages of websites.

We further evaluate the Precision, Recall, and F1-score of \ours in identifying early-stage traffic.
Table~\ref{tab:closed-single-tab} presents a comparison of \ours with existing WF attacks. \ours significantly outperforms other attacks in all stages of page loading. 
For instance, when websites are loaded to 20\%, 30\%, 40\%, 50\%, and 60\%, the F1-score of \ours shows an average increase of 330.43\%, 245.52\%, 151.51\%, 79.59\%, and 38.85\% over existing attacks, respectively. 
For early-stage traffic, we observe that \ours exhibits higher Precision than Recall. This indicates that \ours is effective in avoiding the misidentification of traffic with insufficient website information. 
Benefiting from the temporal distribution analysis of website features and website-adaptive data augmentation, \ours is capable of effectively identifying early-stage traffic that contains sufficient website information while avoiding misidentification of early-stage traffic without adequate website information.

\subsection{Open-World Evaluation}
\begin{figure}[t]
  \centering
  \includegraphics[width=\linewidth]{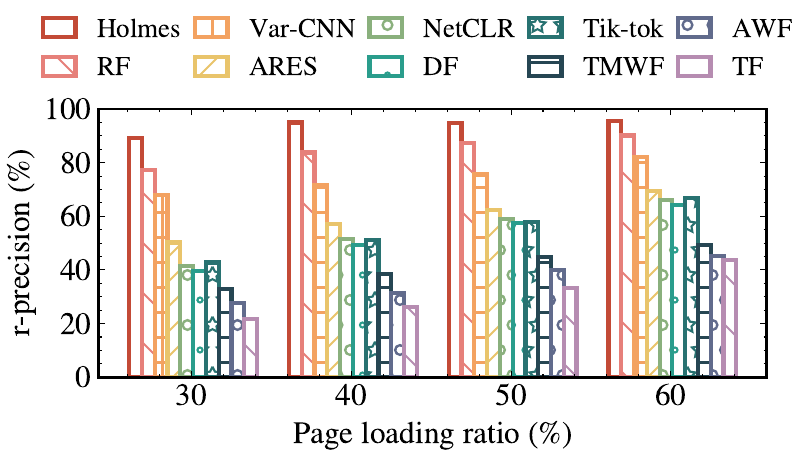}
  \caption{\xinhao{Comparison of the \texttt{r-precision} of WF attacks for early-stage traffic in the open-world scenario.}}
  \label{fig:ow_early_f1}
\end{figure}

We further evaluate the realistic open-world scenario using the dataset of Alexa-top websites, including 95 monitored websites and 40,000 unmonitored websites.
The number of unmonitored websites significantly exceeds the number of monitored websites.
To effectively assess attack performance in the open-world setting, we follow previous works~\cite{wang2020high} by utilizing \texttt{r-precision} for evaluation.

Figure~\ref{fig:ow_early_f1} shows the comparison of \texttt{r-precision} for WF attacks when the ratio of page loading ranges from 30\% to 60\%. 
\ours consistently achieves high \texttt{r-precision} across different page loading ratios.
Compared to existing attacks, \ours demonstrates a significant advantage in identifying early-stage traffic in the open-world scenario. 
For example, when the ratio of page loading is 40\%, \ours achieves the \texttt{r-precision} of 94.96\%, 
while the F1-scores for \rf, \varcnn, \ares, \netclr, \df, \tiktok, \tmwf, \awf, and \tf are 83.77\%, 71.72\%, 56.99\%, 51.49\%, 49.33\%, 51.26\%, 38.46\%, 31.31\%, and 26.06\%, respectively. 
When websites are loaded to 30\%, 40\%, 50\%, and 60\%, the \texttt{r-precision} of \ours shows an average increase of 130.61\%, 109.91\%, 79.00\%, and 57.62\% over existing attacks, respectively.

The experimental results demonstrate that \ours can effectively distinguish between early-stage traffic from monitored and unmonitored websites in the open-world scenario.
Particularly, \ours reduces training overhead compared to baselines by eliminating the requirement for training samples from unmonitored websites. 
\ours leverages the spatial distribution of monitored websites in the feature space. By comparing the distance of unknown traffic in the feature space to the centroid of the website and the website's radius, \ours achieves early-stage WF attacks with high precision in the open-world scenario.

\subsection{Robustness Evaluation}
\label{sec:robust_evaluation}
\begin{figure*}[t]
  \centering
  \includegraphics[width=\linewidth]{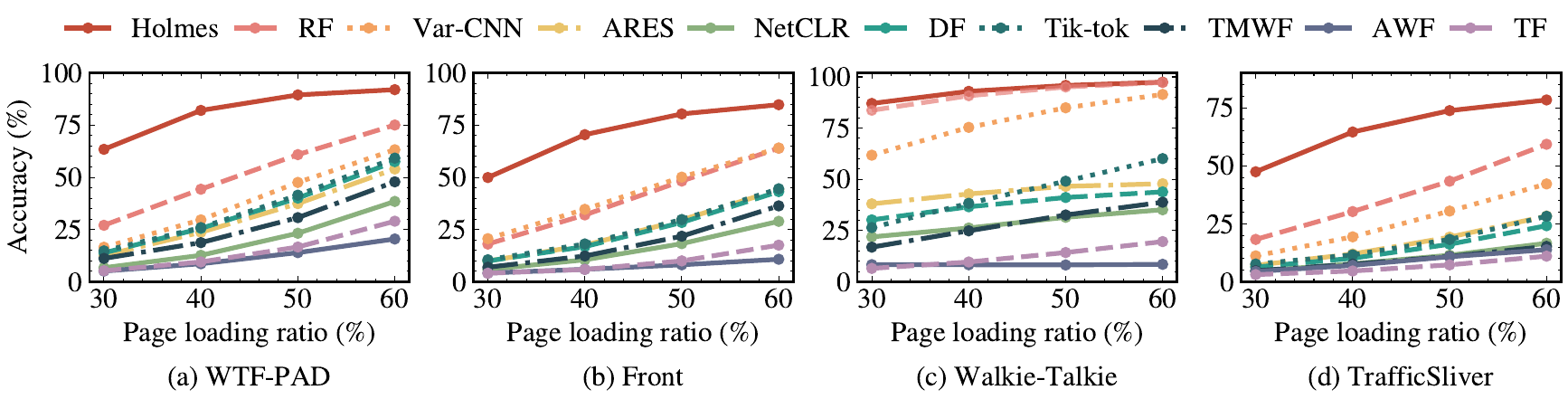}
  \caption{Evaluating robustness of WF attacks for early-stage traffic with four defenses.}
  \label{fig:defense_acc}
\end{figure*}

Next, we evaluate the robustness of \ours using datasets of Alexa-top 95 websites with four defenses.
In Figure~\ref{fig:defense_acc}, we demonstrate the accuracy of WF attacks in different loading stages of websites under defenses.

As shown in Figure~\ref{fig:defense_acc}(a), for the WTF-PAD defense, \ours achieves the best accuracy across all ratios of page loading. 
For early-stage traffic, \ours is more robust compared to other attacks. 
When the page loading ratio is 40\%, \ours achieves an accuracy of 82.03\%, while the accuracy of all baselines is below 45\%. 
For early-stage traffic when websites are 50\% loaded, \ours achieves an accuracy of 89.45\%, 
marking significant improvements over \rf, \varcnn, \ares, \netclr, \df, \tiktok, \tmwf, \awf, and \tf by 46.95\%, 88.04\%, 138.98\%, 284.73\%, 123.23\%, 115.65\%, 191.18\%, 539.84\%, and 436.59\%, respectively.
Similar to \ours, \netclr and \tf generate embeddings of traffic features based on contrastive learning and metric learning, respectively.
However, the accuracies of \netclr and \tf for early-stage traffic with WTF-PAD defense are both below 30\%. 
The advantage of \ours is attributed to its feature extraction and SCL-based traffic embedding, which enable robust website identification under defenses.

Front is a more powerful padding-based defense compared to WTF-PAD. 
By padding dummy packets at the front of the traffic, Front significantly impacts the identification of early-stage traffic.
Figure~\ref{fig:defense_acc}(b) shows the evaluation of WF attacks under Front defense. \ours achieves the best accuracy across all page loading ratios. 
When the page loading ratio is 30\%, 40\%, 50\%, and 60\%, \ours improves the accuracy of baselines by 561.40\%, 480.92\%, 316.03\%, and 192.97\% on average, respectively. 
Existing WF attacks rely on the complete features of individual traffic, whereas \ours leverages the correlation between early-stage traffic and complete traffic of the same website to achieve a more robust WF attack.

In Figure~\ref{fig:defense_acc}(c), we show the accuracy of WF attacks under the Walkie-Talkie defense.
We find that the attack performance of \rf is close to that of \ours. 
The reason is that traffic aggregation information based on time windows has been proven to effectively undermine the Walkie-Talkie defense~\cite{shen2023rf}. 
\ours still holds an advantage in identifying early-stage traffic. 
For instance, at a page loading ratio of 30\%, \ours achieves an accuracy of 87.04\%, 
while the accuracy of \rf, \varcnn, \ares, \netclr, \df, \tiktok, \tmwf, \awf, and \tf are 83.70\%, 61.80\%, 38.07\%, 21.93\%, 30.24\%, 26.47\%, 16.93\%, 8.39\%, and 6.56\%, respectively.

TrafficSliver is a potent defense that combats WF attacks by splitting traffic. 
Figure~\ref{fig:defense_acc}(d) shows the comparison of WF attacks under TrafficSliver defense. 
We observe a significant decrease in the accuracy of baselines under TrafficSliver defense, while \ours maintains its robustness. 
When the page loading ratios are 30\%, 40\%, 50\%, and 60\%, the accuracy of \ours is improved by an average of 711.28\%, 593.93\%, 417.82\%, and 283.98\% compared to other WF attacks.
TrafficSliver is effective in reducing the amount of website information in the early-stage traffic. However, TrafficSliver cannot disrupt the correlation between traffic from different stages of page loading. Therefore, \ours is more robust against the TrafficSliver defense compared to baselines.

\subsection{Reliability Evaluation}
\label{sec:eval-reliability}

\begin{figure}[t]
  \centering
  \includegraphics[width=\linewidth]{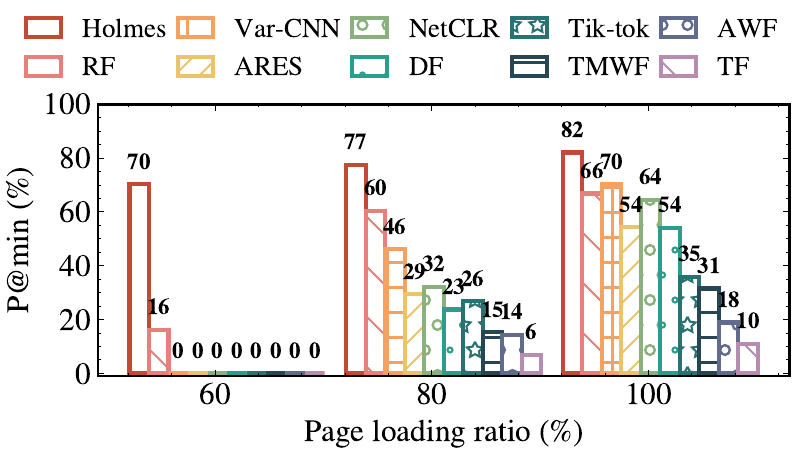}
  \caption{Reliability evaluation of WF attacks under WTF-PAD defense, where \pmin is the minimum of identification Precision for all websites.}
  \label{fig:reliable_pmin}
\end{figure}


The page loading speeds vary significantly across different websites, making it difficult to ensure high precision in detecting early-stage traffic of all websites.
Therefore, we use the minimum precision among all websites (i.e., \pmin) to evaluate the reliability of WF attacks on early-stage traffic. 

Figure~\ref{fig:reliable_pmin} illustrates the reliability of WF attacks under WTF-PAD defense. 
We use the dataset of Alexa-top 95 websites with WTF-PAD defense for evaluation because the variation of WTF-PAD defense based on circuit-level padding has been practically deployed in Tor~\cite{circuitpadding}.
When the page loading ratio is 60\%, \ours achieves the best \pmin of 70.25\%,
while \varcnn, \ares, \netclr, \df, \tiktok, \tmwf, \awf, and \tf have the \pmin of 0. 
The \pmin equals 0 means there are websites that these WF attacks cannot identify.
For traffic at page loading rates of 80\% and 100\%, \ours achieves an average \pmin improvement of 299.43\% and 160.60\% over baselines, respectively.
We find that multi-tab WF attacks, \ares and \tmwf, fail to ensure reliable identification under obfuscated traffic. Existing attacks focus only on high average precision, ignoring the low \pmin caused by differences between websites.
Particularly, for traffic during the complete loading of websites, the reliability of existing WF attacks is limited.
\rf, \tiktok, \ares, and \df, which claim to be robust attacks capable of undermining WTF-PAD defense, achieve high average precisions of 96.78\%, 94.51\%, 91.09\%, and 91.19\% in our evaluation.
However, the \pmin for \rf, \tiktok, \ares, and \df are only 66.87\%, 64.46\%, 54.30\% and 54.11\%, respectively. 
In contrast, \ours significantly improves the reliability of WF attacks and achieves the best \pmin of 82.11\%.

The reliability of \ours is attributed to three aspects:
\first \ours achieves adaptive data augmentation based on the unique temporal distribution of each website, ensuring high precision in the identification of early-stage traffic across all websites.
\second \ours employs supervised contrastive learning to transform features, effectively separating traffic from different websites in the new feature space, thus reducing the misclassification of similar websites. 
\third \ours calculates the spatial distribution features of website traffic in the feature space and enhances the reliability of identification by assessing the correlation between unknown traffic and the unique spatial distribution of each website.

\newcommand{\redtriangle}{%
    \tikz \fill[red] (0,0) -- (0.15cm,0) -- (0.075cm,0.15cm) -- cycle;%
}

\newcommand{\bluetriangle}{%
    \tikz \fill[blue] (0,0.15cm) -- (0.15cm,0.15cm) -- (0.075cm,0) -- cycle;%
}

\begin{table}[t]
\small
\centering
\caption{{Comparison with existing attacks using the dataset of dark web websites in real-world evaluation.}} 
\label{tab:darkweb-time}
\begin{minipage}{\linewidth}
\centering
\begin{tabular}{c|ccc}  \toprule
\textbf{Attacks} & \textbf{Latency} \textsuperscript{\bluetriangle} \footnote{\bluetriangle ~indicates lower is better and \redtriangle ~indicates higher is better.} & \textbf{Loading ratio} \textsuperscript{\bluetriangle} & \textbf{Precision} \textsuperscript{\redtriangle} \\ \midrule 

\tf     &  162.44 s  &  73.67\%  &  17.14  \\
\awf    &  100.91 s  &  47.15\%  &  10.18  \\
\tmwf   &  236.58 s  &  97.58\%  &  47.21  \\
\tiktok &  162.21 s  &  73.63\%  &  63.19  \\
\df     &  162.21 s  &  73.63\%  &  33.68  \\
\netclr &  162.20 s  &  73.61\%  &  28.45  \\
\ares   &  236.58 s  &  97.58\%  &  52.09  \\
\varcnn &  162.21 s  &  73.63\%  &  67.31 \\
\rf     &  162.21 s  &  73.63\%  &  84.99 \\
{$\textsf{RF}_\mathtt{30\%}$}  &  52.44 s  &  25.07\%  &  83.70 \\
\textbf{\ours} & \textbf{45.25 s} & \textbf{21.71\%} & \textbf{85.19} \\
\bottomrule
\end{tabular}
\end{minipage}
\end{table}

\subsection{Real-World Evaluation}

Next, we evaluate \ours using the dataset of 80 dark web websites collected from the real world.
Alexa-top websites are widely used for evaluating WF attacks~\cite{sirinam2018deep, rimmer2018automated, shen2023rf, deng2023robust, sirinam2019triplet}. However, the ranking of Alexa-top websites is based on the interests of all internet users, which may not accurately represent the interests of Tor users in the real world.
Based on the measurements of Tor onion services~\cite{wang2023comprehensive}, we selected 80 of the most popular dark web websites.
We utilized 20 servers deployed across three different countries to collect dark web traffic in August 2023 and April 2024. 
Therefore, this dataset encompasses traffic under various network conditions and traffic exhibiting concept drift due to changes in the websites.
We replay packets of testing traffic to evaluate the time overhead and performance of different WF attacks. 
Moreover, the adversary cannot know the end time of the page loading in advance. 
We set up baselines to end traffic collection when the number of packets meets the input requirements or when no new packets are collected within 1 second. 
Particularly, we use one NVIDIA GeForce RTX 4090 to accelerate the inference of DL models.

Table~\ref{tab:darkweb-time} shows the comparison of WF attacks using the dataset of dark web websites in the real-world evaluation.
The attack latency refers to the average time taken to collect and identify unknown traffic, while the loading ratio represents the average page loading ratio when the website identification result is obtained from the WF attack. 
In particular, we optimize the best-performing attack \rf. $\textsf{RF}_\mathtt{30\%}$ represents \rf attacks with the packet sequence lengths reduced to 30\% of the original length.
We adjust the input sequence lengths of the \rf and retrain the models.
Reducing the input length significantly optimizes latency, but also compromises the identification precision of \rf.
We find that compared to existing attacks and enhanced \rf, \ours exhibits the best attack efficiency and identification precision.
Specifically, \ours reduces latency by an average of 66.33\% and improves precision by an average of 169.36\% compared to baselines.

Dark web websites utilize onion services for server anonymization, requiring more Tor relays and additional time overhead to load. 
We additionally use the dataset of Alexa-top websites under WTF-PAD defense for real-world evaluation. \ours outperforms existing attacks and enhanced \rf in terms of latency and performance.
For Alexa-top websites, \ours reduces latency by an average of 66.38\% and increases precision by an average of 32.32\% compared to baselines.
\ours's advantages are attributed to adaptive data augmentation for different websites and leveraging the spatial distribution of websites for adaptive traffic collection and high-precision website identification. 

\begin{figure}[t]
  \centering
  \includegraphics[width=0.9\linewidth]{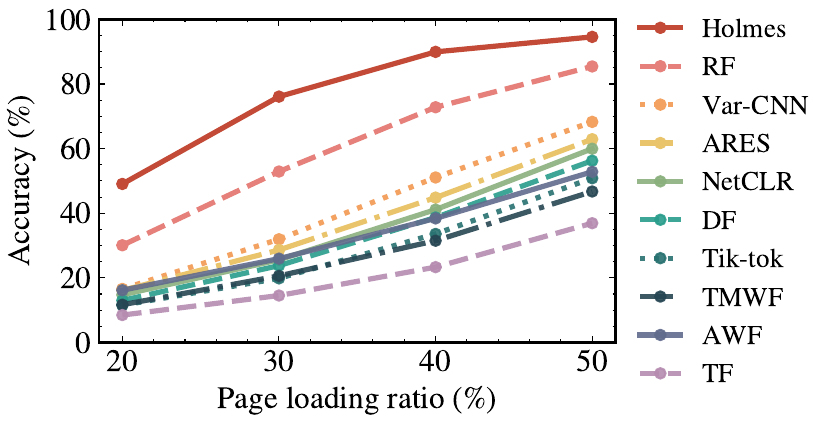}
  \caption{\xinhao{Comparison with enhanced baselines with the early-stage traffic of Alexa-top websites.}}
  \label{fig:random_mask}
\end{figure}

\subsection{Comparison with Enhanced Baselines}

In this section, we enhance the baselines and compare them with \ours. 
Similar to the data augmentation module of \ours, we generate early-stage traffic by masking the tail of the traffic with random lengths, which is added to the training datasets of baselines. 
We evaluate the accuracy of WF attacks on early-stage traffic using the dataset of the Alexa-top 95 websites. 
As shown in Figure~\ref{fig:random_mask}, \ours maintains a significant advantage in identifying early-stage traffic compared to the enhanced baselines. 
When the page loading ratio is 20\%, 30\%, 40\%, and 50\%, \ours' accuracy improved by an average of 255.78\%, 215.09\%, 136.12\%, and 72.15\% compared to the enhanced baselines.

\ours utilizes the temporal distribution of websites to achieve website-adaptive data augmentation, effectively generating early-stage traffic that contains sufficient website information. 
Furthermore, \ours employs supervised contrastive learning to extract the correlations between early-stage traffic and complete traffic, which enables more effective correlation analysis between samples compared to traditional supervised learning.

\begin{figure*}[t]
  \centering
  \includegraphics[width=0.95\linewidth]{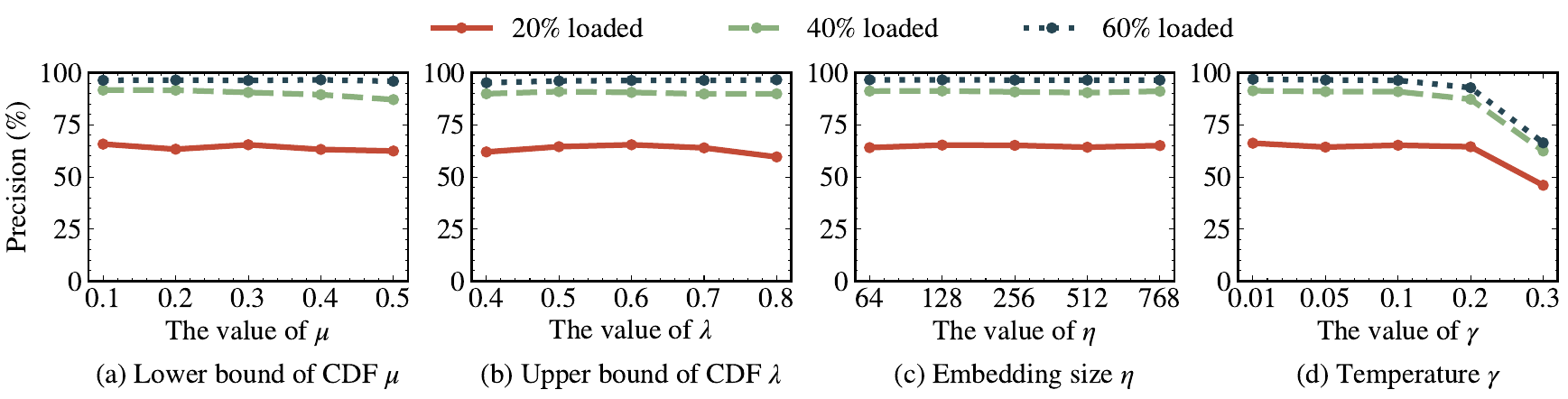}
  \caption{Evaluation of \ours with different parameter settings. We show the identification precision of \ours at 20\%, 40\%, and 60\% loading stages, respectively.}
  \label{fig:parameter_analysis}
\end{figure*}

\subsection{Parameters Analysis}

We further study the impact of different parameter values on the performance of \ours. 
We select four key parameters, including the lower bound of the cumulative time distribution $\mu$, the upper bound of the cumulative time distribution $\lambda$, the embedding size $\eta$, and the temperature $\gamma$. 
We measure the accuracy of \ours when the website loading ratios are 20\%, 40\%, and 60\%, respectively.

As shown in Figure~\ref{fig:parameter_analysis}, we show the accuracy of \ours under different parameter settings. The performance of \ours is insensitive to the settings of the lower bound $\mu$, upper bound $\lambda$, and embedding size $\eta$. 
For example, when the embedding size $\eta$ is increased from 64 to 768, the accuracy of \ours for the traffic of 20\% loaded ranges from 64.06\% to 65.29\%. For the traffic of 60\% loaded, the accuracy of \ours ranges from 96.45\% to 96.58\%. 
Moreover, we observe that a larger temperature $\gamma$ leads to a decrease in the performance of \ours. The reason is that larger temperature $\gamma$ will make model training difficult.
Particularly, the performance of \ours is still stable when the temperature $\gamma$ is less than 0.15. 
In general, the performance of \ours is not sensitive to parameter choices.

\section{Discussion}
\label{sec:discussion}

\noindent \textbf{Concept Drift.} 
The changing content of websites over time can lead to a decline in the effectiveness of WF attacks, i.e., concept drift. 
Concept drift can be addressed by periodically collecting new traffic and retraining models~\cite{rimmer2018automated,deng2023robust,sirinam2019triplet, bahramali2023realistic}. 
However, there are two key challenges. 
\first Detecting concept drift is difficult, and existing attacks detect concept drift by observing the degradation of attack performance.
\second Collecting traffic from all websites and retraining models is time-consuming and resource-intensive.
\ours can effectively detect concept drift samples for each website in the open-world setting. 
Furthermore, \ours does not require frequent model retraining. 
We only collect traffic from websites with concept drift and update the centroid and radius of the corresponding websites. 

\noindent \textbf{Multi-tab Browsing.} 
In recent years, the identification of obfuscated traffic in multi-tab browsing has been widely studied~\cite{deng2023robust, jin2023transformer}.
In fact, multi-tab WF attacks can be transformed into multiple single-tab WF attacks.
The adversary at the guard node can split obfuscated traffic based on the circuit ID~\cite{sirinam2018deep,sirinam2019triplet}.
On the other hand, \ours can be used to enhance the performance of existing multi-tab attacks and trained used obfuscated traffic under multi-tab settings. 
For instance, \ours can replace the Trans-WF model in the multi-tab attack framework ARES~\cite{deng2023robust}, effectively identifying websites in the early stages of page loading.

\noindent \textbf{Countermeasure against \ours.} 
\ours exploits the temporal and spatial distribution of website traffic. The spatial distribution can be disrupted through traffic obfuscation. 
One possible design is as follows.
The Defender collects traffic in advance to calculate the spatial distribution of website traffic. 
Then the defender utilizes GAN to generate obfuscated traffic based on the spatial distribution of website traffic so that the distance between the obfuscated traffic and the centroid of the website increases. 
We leave an in-depth exploration of this design to future work.

\noindent \textbf{Limitations of \ours.}
First, \ours may not be able to accurately identify websites with the same template and similar content because they generate similar traffic patterns. 
Second, Tor software updates and significant modifications of website content lead to changes in website traffic patterns, which may impact the performance of \ours. 
We aim to further improve the practicality of WF attacks in future work.

\section{Related Work}

\noindent \textbf{DL-based WF Attacks.} Recently deep learning has been widely applied to construct website fingerprinting attacks~\cite{rimmer2018automated, sirinam2018deep, rahman2019tik, bhat2019varcnn, shen2023rf, deng2023robust, jin2023transformer, bahramali2023realistic}. 
DL-based WF attacks demonstrate outstanding attack performance. 
However, these attacks require traffic close to the completion of page loading to identify websites.
\ours leverages the temporal distribution and spatial distribution of website traffic, enabling the extraction of correlations among website traffic. 
Therefore, our constructed attack can achieve robust and reliable WF attacks based on the early-stage traffic of page loading.

\noindent \textbf{Practical WF Attacks.} 
The feasibility of deploying existing WF attacks in the real world is hampered by strong assumptions~\cite{cherubin2022online, juarez2014critical}.
Recent works aim to relax these assumptions in real-world settings, e.g., multi-tab browsing~\cite{jin2023transformer, deng2023robust}, robust WF attacks against defenses~\cite{shen2023rf}, attacks with a small number of training samples~\cite{sirinam2019triplet}, dynamic network conditions~\cite{bahramali2023realistic}, open-world attacks~\cite{wang2020high}.
\ours aims to accurately identify websites at a very early stage of page loading, further enhancing the practicality of WF attacks.

\noindent \textbf{Early-Stage Traffic Analysis.} Early-stage traffic analysis facilitates real-time processing of traffic, which is crucial for throttling malicious traffic~\cite{chuanpu2024ccs, qing2023low, li2022dynamic, deng2024cache}. 
Most existing studies focus on early-stage non-encrypted traffic analysis, where traffic can be accurately identified by using a small number of packets~\cite{huang2008early, gomez2009early}.
The challenge intensifies if the traffic under analysis is encrypted~\cite{qu2012accuracy}.
Recently, DL-based traffic analysis methods~\cite{zhan2021website, wang2022two} achieve accurate early-stage encrypted traffic classification in specific scenarios.
However, existing methods cannot achieve WF attacks in the early stages under Tor traffic.
\ours achieves early-stage WF attacks by analyzing the spatial-temporal correlations among website traffic.
To the best of our knowledge, \ours is the first early-stage traffic analysis for Tor traffic.
\section{Conclusion}
\label{sec:conclusion}

In this paper, we propose \ours, a reliable and robust early-stage WF attack. 
Specifically, \ours utilizes the temporal distribution of website traffic to achieve website-adaptive data augmentation and employs supervised contrastive learning to embed traffic into a low-dimensional feature space. \ours calculates the correlation of early-stage traffic with each website by leveraging the spatial distribution of website traffic in the embedding space, thereby enabling early-stage website identification. 
We conduct extensive evaluations of \ours using six datasets, and the experiment results demonstrate its effectiveness in identifying early-stage traffic.

\section*{Acknowledgment}

We thank our anonymous reviewers for their helpful comments and feedback. 
The work is supported in part by NSFC
under Grant 62132011 and 62425201.
Qi Li is the corresponding author of this paper.

\bibliographystyle{ACM-Reference-Format}
\balance
\bibliography{references}

\end{document}